\documentclass[11pt]{article}
\usepackage{graphicx}
\usepackage{cite}
\usepackage{amsthm}
\usepackage{amsmath}
\usepackage{amssymb}

\def \d {{\rm d}}
\def \U {{\mathcal U}}
\def \V {{\mathcal V}}
\newcommand{\Up}{U_{+}}
\newcommand{\ds}{{\mbox{\b{d}}}}
\newcommand{\bZ}{\bar{Z}}

\newcommand{\bzeta}{\bar{\zeta}}
\newcommand{\im}{{\rm{i}\,}}

\begin{document}

\title{Penrose junction conditions extended:  \\
       impulsive waves with gyratons}

\author{
J.~Podolsk\'y$^1$\thanks{{\tt podolsky@mbox.troja.mff.cuni.cz}},\,\,
R.~\v{S}varc$^1$\thanks{{\tt robert.svarc@mff.cuni.cz}},\,\,
R.~Steinbauer$^2$\thanks{{\tt roland.steinbauer@univie.ac.at}}\,\,
and
C.~S\"amann$^2$\thanks{{\tt clemens.saemann@univie.ac.at}}
\\ \\ \\
$^1$ Institute of Theoretical Physics,\\
Charles University, Faculty of Mathematics and Physics, Prague \\
V Hole\v{s}ovi\v{c}k\'ach 2, 18000 Prague 8, Czech Republic.\\ \\ \\
$^2$ Faculty of Mathematics, University of Vienna, \\
Oskar-Morgenstern-Platz 1, 1090 Vienna, Austria. \\ \\
}
\date{\today}

% 2014 - April 2017

% PACS
% 04.20.Jb Exact solutions
% 04.30.-w Gravitational waves: theory
% 04.30.Nk Wave propagation and interactions

%MSC
%83C15   	Exact solutions
%83C35   	Gravitational waves
%83C10   	Equations of motion

%Keywords: impulsive gravitational waves, junction conditions, geodesics, nonexpanding impulses, continuous metric form

\maketitle

\vspace{3mm}

\begin{abstract}
We generalize the classical junction conditions for constructing impulsive gravitational waves by the  Penrose ``cut and paste'' method. Specifically, we study nonexpanding impulses which propagate in  spaces of constant curvature with any value of the cosmological constant (that is Minkowski, de Sitter, or anti-de~Sitter universes) when additional off-diagonal metric components are present. Such components encode a possible angular momentum of the ultra-relativistic source of the impulsive wave --- the so called gyraton. We explicitly derive and analyze a specific transformation that relates the distributional form of the metric to a new form which is (Lipschitz) continuous. Such a transformation automatically implies an extended version of the Penrose junction conditions. It turns out that the conditions for identifying points of the background spacetime across the impulse are the same as in the original Penrose ``cut and paste'' construction, but their derivatives now directly represent the influence of the gyraton on the axial motion of test particles. Our results apply both for vacuum and nonvacuum solutions of Einstein's field equations, and can also be extended to other theories of gravity.
\end{abstract}

\vspace{5mm}

\section{Introduction}

An impulsive gravitational wave can most intuitively be understood as a limit of a suitable family of sandwich waves with their profiles approaching the Dirac delta. Formally, if $d_\varepsilon(\U)$ denotes the sandwich wave profile where $\U$ is the retarded time coordinate, it is assumed that  the supports of the sequence $d_\varepsilon(\U)$ shrink to zero as ${\varepsilon\to0}$ (the sandwich waves have ``ever shorter  duration'' $\varepsilon$) but simultaneously their amplitudes become bigger (the waves are ``ever stronger'' as ${\varepsilon^{-1}}$). Such a distributional limit ${d_\varepsilon(\U)\to \delta(\U)}$ gives a solution, which (at least formally) represents an impulsive wave localized on a \emph{single wave-front}~${\U=0}$.

In the simplest yet important context of vacuum {\it pp\,}-waves propagating in Minkowski space, this procedure was first explicitly considered in \cite{Pen68a,Pen68b,Pen72,Rindler} and later elsewhere (see, e.g., \cite{[A3]}). Employing the well-know Brinkmann form
of {\it pp\,}-waves \cite{KSMH,GP:2009}, one directly obtains the metric
\begin{equation}
  \d s^2= 2\,\d\eta\,\d\bar\eta-2\,\d \U\,\d \V +2H(\eta,\bar\eta)\,\delta(\U)\,\d \U^2\,.
\label{imp-pp}
\end{equation}
Analogously, more general nonexpanding planar impulsive waves of the Kundt type and expanding spherical impulsive waves of the Robinson--Trautman type can be constructed. A nonvanishing cosmological constant $\Lambda$ can also be considered, so that the impulsive waves may propagate in any space of constant curvature --- that is Minkowski, de Sitter, or anti-de~Sitter
universe. Detailed accounts of these spectimes, various methods of their construction, their mutual relations, the most important examples, and a number of references can be found in \cite{Pod2002b,BarHog2003,GP:2009} and in the review parts of the recent works \cite{PSSS:2015, PSSS:2016}.

However, although the classical metric (\ref{imp-pp}) has been employed and investigated in many works, it is \emph{not} the most general form of impulsive {\it pp\,}-waves. As considered already in the original paper by Brinkmann \cite{Brinkmann:1925}, additional \emph{off-diagonal terms can be added},\footnote{In fact, the most general Brinkmann geometry also admits higher dimensions and the possibility that the transverse Riemannian space is not flat, see \cite{CandelaFloresSanchez:2003, FloresSanchez:2006, PodolskyZofka:2009, PodolskySvarc:2012, PodolskySvarc:2013a, SSS:2016}.} leading to the metric
\begin{eqnarray}
  \d s^2 \!\!\!&=&\!\!\! 2\,\d\eta\,\d\bar\eta-2\,\d \U\,\d \V  +2H(\eta,\bar\eta)\,\delta(\U)\,\d \U^2   \nonumber\\
  &&\!\!\!   +2J(\eta,\bar\eta,\U)\,\d\eta\,\d \U + 2\bar J(\eta,\bar\eta,\U)\,\d\bar\eta\,\d \U  \,.
\label{imp-pp_gyrat}
\end{eqnarray}
In \emph{vacuum regions}, it is a standard and common approach to completely remove these additional metric terms by a suitable coordinate transformation. However, such a gauge freedom is \emph{only local} and ignores the global (topological) properties of the spacetimes. By neglecting the metric function $J$ in (\ref{imp-pp_gyrat}), an important physical property of the spacetime is eliminated, namely the possible rotational character of the source of the gravitational wave (its internal spin/helicity).

This remarkable fact was first noticed by Bonnor \cite{Bonnor:1970b,Griffiths:1972}, see \cite{GP:2009}, who studied both the interior and the exterior field of a ``spinning null fluid'' in the class of axially symmetric \emph{pp}-waves. Spacetimes with such a localized spinning source moving at the speed of light, whose angular momentum is encoded in the function $J$, were independently rediscovered in 2005 \cite{FrolovFursaev:2005,FrolovIsraelZelnikov:2005,FrolovZelnikov:2005,FrolovZelnikov:2006,YoshinoZelnikovFrolov:2007}, their physical application as a model of relativistic particles was emphasized, and they were called ``gyratons''. These \emph{pp}-wave-type gyratons were then investigated in greater detail and also generalized to higher dimensions and various nonflat backgrounds in a wider Kundt class which may also include a cosmological constant or an additional  electromagnetic field, see
\cite{KadlecovaZelnikovKrtousPodolsky:2009,KadlecovaKrtous:2010,KrtousPodolskyZelnikovKadlecova:2012,PodolskySteinbauerSvarc:2014} for more details and further references.

It is well known that there exist several distinct methods of constructing impulsive {\it pp\,}-waves represented by the metric (\ref{imp-pp}), namely the ``cut and paste'' method with Penrose junction conditions, explicit construction of continuous coordinates, distributional limits of sandwich waves, boosts of specific initially static sources, and embedding from higher dimensions. These were recently reviewed, e.g., in \cite{GP:2009,Pod2002b,BarHog2003,PSSS:2015}. Surprisingly, apart from the straightforward distributional limit of sandwich waves, the other construction methods have not yet been given for the more general impulsive metric (\ref{imp-pp_gyrat}) with gyratonic terms given by nonzero $J$. It is now our goal in this paper to derive such a generalization, that is to find generalized Penrose junction conditions in his ``cut and paste'' method, and to discover the corresponding continuous metric form of impulsive {\it pp\,}-waves with gyratons. We will also find its relation to the distributional metric form (\ref{imp-pp_gyrat}).

First, however, it is necessary to briefly summarize the main construction methods for the classical (non-gyratonic) impulsive {\it pp\,}-waves (\ref{imp-pp}).

\subsection{Penrose junction conditions and the ``cut and paste'' method}
\label{subs1}

\begin{figure}[htp]
\centerline{\includegraphics[scale=0.5]{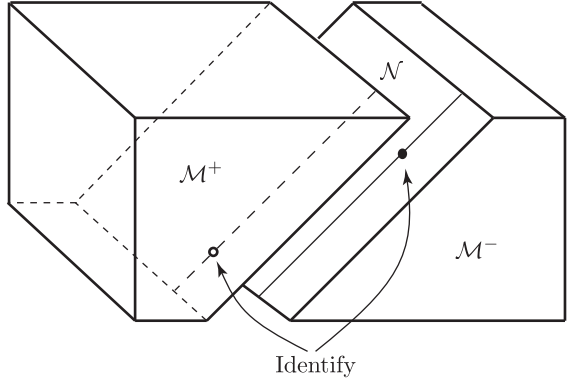} }
\caption{ \small Minkowski space is cut into two parts ${\cal M}^-$ and ${\cal M}^+$ along a plane null hypersurface~${\cal N}$. These parts are then ``re-attached'' with an arbitrary ``warp'' in which points are shunted along the null generators of the cut and then identified. This generates an impulsive gravitational wave in flat space.}
\label{planecut}
\end{figure}

Penrose in \cite{Pen68a, Pen68b} and in a seminal work \cite{Pen72} presented a geometrical ``cut and paste'' method for constructing a class of impulsive waves in flat background, represented by the metric (\ref{imp-pp}). It is based on  cutting Minkowski space ${\cal M}$ along a plane null hypersurface ${\cal N}$ and then ``re-attaching'' the two ``halves'' ${\cal M}^-$ and ${\cal M}^+$ by identification of boundary points with a specific ``warp'', see Figure~\ref{planecut}.

More explicitly, the Penrose method first removes (by a ``cut'') the plane null hypersurface ${\cal N}$ given by
${\U=0}$ from flat spacetime in the form
 \begin{equation}
\d s_0^2= 2\,\d\eta\,\d\bar\eta -2\,\d\U\,\d\V  \,,
 \label{Mink}
 \end{equation}
and then re-attaches (by a ``paste'') the halves ${{\cal M}^-(\U<0)}$ and ${{\cal M}^+(\U>0)}$
by making the identification of boundary points with a ``warp'' in the coordinate $\V$ such that
\begin{equation}
\Big[\,\eta,\,\bar\eta,\,\V,\,\U=0_-\,\Big]_{_{{\cal M}^-}}\equiv
\Big[\,\eta,\,\bar\eta,\,\V-H(\eta,\bar\eta),\,\U=0_+\,\Big]_{_{{\cal M}^+}}\ , \label{junctions}
\end{equation}
where $H(\eta,\bar\eta)$ is \emph{any} real-valued function of $\eta$ and $\bar\eta$. It was shown in \cite{Pen72} that these \emph{Penrose junction conditions} (\ref{junctions}) automatically guarantee that the Einstein field equations are satisfied everywhere \emph{including} on ${\,\U=0}$. Thus gravitational (plus possibly null-matter) impulsive waves are obtained.

\subsection{Continuous coordinates for impulsive {\it pp\,}-waves}

The Penrose ``cut and paste'' approach is an elegant and general method because, by prescribing the junction conditions
(\ref{junctions}) in  (\ref{Mink}), all impulsive gravitational waves in Minkowski space can be constructed. However, the formal identification of points on both sides of the impulsive hypersurface does not directly yield explicit metric forms of the entire spacetimes.

It is thus crucial to know a suitable coordinate system for these solutions in which the metric is explicitly continuous everywhere, including on the impulse. Such a metric reads
 \begin{equation}
\d s^2= 2\,|\d Z+\Up(H_{,\bar Z Z}\d Z+H_{,\bar Z\bar Z}\d\bar Z)|^2-2\,\d U\d V\,,
 \label{contipp}
 \end{equation}
where $H(Z,\bar Z)$ is an \emph{arbitrary} real-valued function while
\begin{align}
\Up\equiv\Up(U) &=
    \begin{cases}
     0 & \text{if } U \leq 0\,, \\
     U & \text{if } U \geq 0
    \end{cases}
\end{align}
is the \emph{kink-function}. Notice that formally ${\Up=U\Theta}$, where ${\Theta=\Theta(U)}$ is the \emph{Heaviside step function}. Since the kink function is Lipschitz continuous, the metric (\ref{contipp}) is locally Lipschitz in the variable $U$ even across the null hypersurface ${U=0}$.\footnote{But $H$ as a function of ${Z, \bar Z}$ may have singularities.}  This implies that the curvature is a distribution (we are within the ``maximal'' distributional curvature framework as identified by Geroch and Traschen \cite{GT:87}). Indeed, the discontinuity in the derivatives of the metric introduces  impulsive components in the Weyl and curvature tensors proportional to the Dirac distribution, namely ${\,\Psi_4 = H_{,ZZ} \,\delta(U)\,}$ and ${\,\Phi_{22} = H_{,Z\bar Z} \, \delta(U)}$, see \cite{PodolskyGriffiths:1999a}. The metric (\ref{contipp}) thus explicitly describes  impulsive waves in Minkowski background: it is of the general Rosen form of impulsive {\it pp\,}-waves \cite{Pen72,[B2],Steinb}. Let us note that the continuous coordinate system for the particular Aichelburg--Sexl solution \cite{AichelburgSexl:1971} was found in \cite{DE78}.

In fact, the continuous form of the impulsive metric (\ref{contipp}) can be obtained systematically by a suitable transformation of the flat Minkowski metric (\ref{Mink}). For ${U>0}$  (in ${\cal M}^+$) the transformation
${ \U=U}$, ${\ \V=V+H+UH_{,Z}H_{,\bar Z}}$, ${\ \eta=Z+UH_{,\bar Z}}$ yields
$\ \d s_0^2= 2\,|\d Z+U(H_{,\bar Z Z}\d Z+H_{,\bar Z\bar Z}\d\bar Z)|^2-2\,\d U\d V$.
This can now be combined with the metric (\ref{Mink}) for ${U<0}$  (in ${\cal M}^-$) in which the identity ${\U=U}$, ${\ \V=V}$, ${\ \eta=Z}$ is applied. The combined transformation relating both parts of (\ref{Mink}) to (\ref{contipp}) is thus
 \begin{eqnarray}
\U  \!\!&=&\!\! U\, , \nonumber\\
\V  \!\!&=&\!\! V+\Theta\,H+\Up\,H_{,Z}H_{,\bar Z}\,,  \label{trans}\\
\eta\!\!&=&\!\! Z+\Up\,H_{,\bar Z}\,. \nonumber
 \end{eqnarray}
It is clearly \emph{discontinuous} in the coordinate $\V$ on ${\U=0}$. Now, using the fact that the global coordinates ${U,V,Z}$ give rise to the continuous form of the metric (\ref{contipp}), we obtain from (\ref{trans})  exactly the Penrose junction conditions (\ref{junctions}) for reattaching the two halves of the spacetime ${\cal M}^-$ and ${\cal M}^+$ with the warp ${\V \to \V-H}$. This  procedure is thus an {\it explicit} Penrose's ``cut and paste'' construction of impulsive gravitational {\it pp\,}-waves.

\subsection{Distributional form of impulsive {\it pp\,}-waves}

Interestingly, the distributional form of the impulsive {\it pp\,}-wave spacetimes is obtained from the continuous form of the metric (\ref{contipp}) by applying the \emph{combined} transformation (\ref{trans}), if we consider also the terms which arise from the derivatives of $\Theta(U)$ and $\Up(U)$. Indeed, (\ref{trans}) relates (\ref{contipp}) formally to
 \begin{equation}
\d s^2= 2\,\d\eta\,\d\bar\eta-2\,\d \U\,\d \V +2H(\eta,\bar\eta)\,\delta(\U)\,\d \U^2\,,
 \label{ppimp}
 \end{equation}
which explicitly includes the impulse  located on the wavefront ${\U=0}$. This is a gravitational wave (or an impulse of null matter) in flat spacetime, depending on the specific form of the function $H$. It is just the Brinkmann form of a general impulsive {\it pp\,}-wave (\ref{imp-pp}).

Of course, the discontinuity in the complete transformation (\ref{trans}) which formally relates the continuous and distributional forms of  impulsive solutions  causes some subtle mathematical problems. In fact, to obtain (\ref{ppimp}) one is forced to use the distributional identities $\Theta'=\delta$ and $\Up'=\Theta$, together with the multiplication rules $\Theta^2=\Theta$ and $\Theta\Up=\Up$. It is well known that in general this leads to inconsistencies, see e.g. \cite[Ex.~1.1.1(iv)]{GKOS:01}. However, it was shown in \cite{Steinb} that (\ref{trans}) is in fact  a {\it rigorous} example of a generalized coordinate transformation in the sense of Colombeau's generalized functions. Moreover, it is possible to interpret this change of coordinates as the distributional limit of a family of smooth transformations which is obtained by a general regularization procedure, i.e., by considering the impulse as a limiting case of sandwich waves with an {\it arbitrarily} regularized wave profile. These results put the formal (``physical'') equivalence of  both continuous and distributional forms of impulsive spacetimes on a solid ground. Therefore, the full family of impulsive limits (\ref{imp-pp}) of sandwich {\it pp\,}-waves is indeed equivalent to the distributional form of the solutions (\ref{ppimp}), and consequently to the continuous metric (\ref{contipp}) obtained by the explicit ``cut and paste'' method (\ref{junctions}) in flat background (\ref{Mink}).

\bigskip
Now we will present generalizations of these main methods of construction of impulsive waves to include gyratons. In the next Section~\ref{Sec2} we will concentrate on the family of impulsive {\it pp\,}-waves in Minkowski space. In Section~\ref{Sec3} we then further generalize our results to all nonexpanding impulsive waves in (anti-)de Sitter space with ${\Lambda\not=0}$.

\newpage

\section{Generalization to include gyratons}
\label{Sec2}

In order to find a continuous metric form for impulsive {\it pp\,}-waves with additional gyratonic terms, as represented by the distributional metric (\ref{imp-pp_gyrat}), it is necessary to generalize the transformation (\ref{trans}). We found the following ansatz  which leads to such a generalization: Set
 \begin{eqnarray}
\U  \!\!&=&\!\! U\, , \nonumber\\
\V  \!\!&=&\!\! V+\Theta\,H+\Up\,H_{,Z}H_{,\bar Z}+W\,,  \label{trans_gyraton}\\
\eta\!\!&=&\!\! \big(Z+\Up\,H_{,\bar Z}\big)\,\exp\big(\,\im F\big)\,, \nonumber
 \end{eqnarray}
where ${H=H(Z,\bar Z)}$. The additional functions ${W=W(Z,\bar Z,U)}$ and ${F=F(Z,\bar Z,U)}$ are assumed to be real-valued, and are taken to satisfy the conditions
 \begin{eqnarray}
F_{,U} \!\!\!&=&\!\!\! \frac{\im \bar J}{Z+\Up\,H_{,\bar Z}}\,\exp\big(\!-\im F\big)\,, \label{condit2}\\
W_{,U} \!\!\!&=&\!\!\! -J \bar{J} \,.  \label{condit1}
 \end{eqnarray}
Let us remark that, in fact, the right-hand side of equation (\ref{condit2}) is always \emph{real} for the standard gyratonic metric of the form (\ref{ro-metric}) which does not involve the off-diagonal term ${\d\rho\, \d \U}$. Indeed, as has been shown in \cite{PodolskySteinbauerSvarc:2014}, this is the most reasonable choice to represent the physically relevant quantities in the metric functions.

By substituting these relations into the  metric (\ref{imp-pp_gyrat}), a straightforward calculation leads to
\begin{eqnarray}
\d s^2 \!\!\!&\equiv &\!\!\!  2\,\big| \ds \zeta +\im \zeta\, \ds F \big|^2
 +2\,\big[\big(\delta-\Theta'\big) H + \big((\Up')^2-\Up'\big) H_{,Z} H_{,\bar Z}\big]\d U^2
  \nonumber\\
&&
 \!\!\!  +2\,\big[\big(\Up'-\Theta\big)\ds H + \big(\Up\Up'-\Up\big)\big(H_{,Z}\,\ds H_{,\bar Z}-H_{,\bar Z}\,\ds H_{,Z}\big)  \nonumber\\
&&
\qquad +\im\Up' \big( \zeta H_{,Z} - \bzeta H_{,\bar Z}\big) \ds F-\ds W \,\big]\d U -2\,\d U\d V
  \,, \label{ContGenMetricSim}
\end{eqnarray}
where we have used the convenient shorthand
\begin{equation}\label{zeta}
\zeta \equiv  Z+\Up\,H_{,\bar Z}\,,
\end{equation}
and the symbol ${\ds}$ stands for the \emph{spatial differentials} of the functions $\zeta$ and  $H$, $F$, $W$, that is
\begin{equation}
 \ds \zeta \equiv \d Z + \Up(H_{,\bar Z Z}\d Z+H_{,\bar Z\bar Z}\d \bar Z) \,,
 \label{spatialdifzeta}
\end{equation}
and
\begin{eqnarray}
 \ds H     \!\!\!&\equiv &\!\!\! H_{,Z}\,\d Z+H_{,\bZ}\,\d\bZ\,, \\
 \ds F     \!\!\!&\equiv &\!\!\! F_{,Z}\,\d Z+F_{,\bZ}\,\d\bZ\,, \\
 \ds W     \!\!\!&\equiv &\!\!\! W_{,Z}\,\d Z+W_{,\bZ}\,\d\bZ\,.
\end{eqnarray}

Finally, employing the standard distributional identities
\begin{equation}
\Theta'=\delta\,, \qquad
\Up'=\Theta\,,
\label{distrident1}
\end{equation}
and the multiplication rules
\begin{equation}
\Theta^2=\Theta\,, \qquad
\Theta\Up=\Up  \,, \quad
\label{distrident2}
\end{equation}
the metric (\ref{ContGenMetricSim}) simplifies considerably to
\begin{equation}
\d s^2 = 2\,\big| \ds \zeta +\im \zeta\, \ds F \big|^2
 +2\,\big[\, \im\Theta \big( \zeta H_{,Z} - \bzeta H_{,\bar Z}\big) \ds F-\ds W \,\big]\d U
 -2\,\d U\d V \,. \label{ContGenMetricSimp}
\end{equation}

We immediately observe that the new metric (\ref{ContGenMetricSimp}) reduces to the continuous metric form (\ref{contipp}) without the gyratonic terms: indeed  ${J=0}$ allows for the solutions ${F=0=W}$. Moreover, (\ref{ContGenMetricSimp}) is continuous provided the spatial differentials $\ds F$ and $\ds W$ are continuous functions of $U$, and $\ds F$ is vanishing at ${U=0}$. Then (\ref{ContGenMetricSimp}) is locally Lipschitz, and the transformation (\ref{trans_gyraton}) is formal precisely in the same way as (\ref{trans}): Indeed, the distributional identities (\ref{distrident1}) and the multiplication rules (\ref{distrident2}) which have to be employed to cancel the terms proportional to $\d U^2$, ${\ds H \d U}$ and ${\ds H_{,Z} \d U}$ are precisely the same as when going from (\ref{ppimp}) to (\ref{contipp}), as can explicitly be seen from the metric (\ref{ContGenMetricSim}). The additional terms generated from $F$ and $W$ cancel purely due to the differential conditions (\ref{condit2}) and (\ref{condit1}), and we will comment on regularity issues in the special cases considered below.
In particular, in the fundamental case when $J$ is proportional to a step-function  $\Theta$, it is expected that $F$ and $W$ involve the kink function $\Up$, and they are vanishing for ${U\le0}$ (so that ${\,\U=U, \V=V, \eta=Z}$ for ${U\le0}$).

Thus, we have reached our aim to generalize (\ref{trans}) in a natural way  to impulsive {\it pp\,}-waves with gyratons, provided we find appropriate functions $F$ and $W$ satisfying equations (\ref{condit2}) and (\ref{condit1}) with ${J\not=0}$. For this, of course, the specific gyraton function $J(\eta,\bar\eta,\U)$ must be re-expressed as a function $J(Z,\bar Z,U)$ using (\ref{trans_gyraton}).

In the next subsection we will explicitly show that the key example has exactly these properties, and thus the metric (\ref{ContGenMetricSimp}) will be a \emph{continuous metric representation of the gyratonic impulsive waves} in that case.

\subsection{The key example}
\begin{figure}[h]
\begin{center}
\includegraphics[scale=0.85]{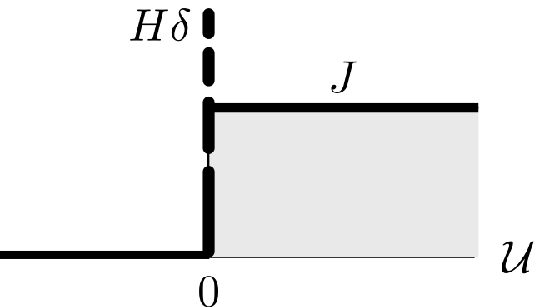}
\hspace{15mm}
\includegraphics[scale=0.83]{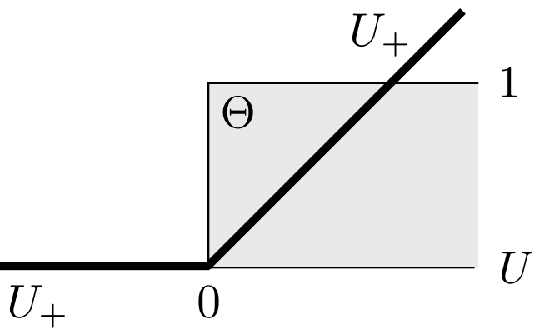}
\end{center}
\caption{ \small Left: An example of the impulsive wave  for which the gyratonic profile $J$ is proportional to the Heaviside step function $\Theta$. Right: Plot of the Heaviside step function $\Theta$ and of the kink function $\Up$~(bold).}
\label{fig2}
\end{figure}

As an illustration, let us now consider the gyratonic extension of standard impulsive {\it pp\,}-waves decribed by the distributional metric (\ref{imp-pp_gyrat}) with the $\d \U^2$-term
\begin{equation}\label{chi_step}
2H(\eta,\bar\eta)\,\delta(\U)\,,\qquad \hbox{and} \quad
J(\eta,\U)=\frac{\chi}{2\im \eta}\, \Theta(\U)\,,
\end{equation}
where $\chi$ is a \emph{constant} (see \cite{FrolovFursaev:2005, YoshinoZelnikovFrolov:2007, PodolskySteinbauerSvarc:2014}) as shown in Figure~\ref{fig2}.
\vspace{0mm}
\vspace{0mm}

\noindent
In such a case, the explicit integration of equations (\ref{condit2}), (\ref{condit1}) with $H(Z,\bZ)$ gives the functions
\begin{align}
  F=&\,\frac{\chi}{2(Z H_{,Z}-\bZ H_{,\bZ})}\,\log\frac{Z\bZ+\Up\,\bZ H_{,\bZ}}{Z\bZ+\Up\,ZH_{,Z}} \,,
\label{OneImpGyrContTrFun_id2}\\
  W=&\,\frac{\chi^2}{4(Z H_{,Z}-\bZ H_{,\bZ})}\,\log\frac{Z\bZ+\Up\,\bZ H_{,\bZ}}{Z\bZ+\Up\,ZH_{,Z}} \,. \label{OneImpGyrContTrFun_id1}
\end{align}
Interestingly, ${W=(\chi/2)\,F}$. It can be seen that both $F$ and $W$ are now indeed real locally Lipschitz continuous functions of~$U$, even on the impulsive surface ${U=0}$. The resulting metric (\ref{ContGenMetricSimp}) becomes
\begin{align}
\d s^2=&\,\, 2\,\big| \d Z +\Up\,\ds(H_{,\bZ}) +\im \big(Z+\Up\,H_{,\bZ}\big)\, \ds F \big|^2 \nonumber\\
 & +2\big[ \im\big(Z\,H_{,Z}-\bZ\,H_{,\bZ}\big)\Theta\,\ds F-\ds W \big]\d U -2\,\d U\d V \,, \label{ContOneImpGyr_id}
\end{align}
where we explicitly have
\begin{align}
\ds F=&\,-\frac{\chi}{2}\log\frac{Z\bZ+\Up\,\bZ H_{,\bZ}}{Z\bZ+\Up\,ZH_{,Z}}\times\frac{\ds(Z H_{,Z})-\ds(\bZ H_{,\bZ})}{(Z H_{,Z}-\bZ H_{,\bZ})^2} \nonumber\\
 &\,+\frac{\chi}{2}
 \frac{\Up\, {\cal H}_1 +\Up^2 \,{\cal H}_2}
 {(Z H_{,Z}-\bZ H_{,\bZ})(Z\bZ+\Up\,ZH_{,Z})(Z\bZ+\Up\,\bZ H_{,\bZ})} \,, \\
\ds W=&\,\frac{\chi}{2}\,\ds F \,,
\label{dif_OneImpGyrContTrFun_id}
\end{align}
with
\begin{align}
{\cal H}_1=&\, (Z H_{,Z}-\bZ H_{,\bZ})(\bZ\d Z -Z\d\bZ) + Z\bZ\,\big( \ds(\bZ H_{,\bZ})-\ds(Z H_{,Z})\big)\,,\\
{\cal H}_2=&\, Z H_{,Z}\,\ds(\bZ H_{,\bZ})-\bZ H_{,\bZ}\,\ds(Z H_{,Z}) \,. \label{ContOneImpGyr_id_new}
\end{align}

Moreover, ${F=0=W}$ \emph{for all} ${U\le0}$ since the $\log$ term in (\ref{OneImpGyrContTrFun_id2}), (\ref{OneImpGyrContTrFun_id1}) vanishes for ${U\le0}$. Hence the \emph{metric} (\ref{ContOneImpGyr_id}) is locally Lipschitz continuous (even across ${U=0}$), and can be written as
\begin{align}
\d s^2=&\,\, 2\,\Big| \d Z +\Up(H_{,\bar Z Z}\d Z+H_{,\bar Z\bar Z}\d \bar Z)
   +\im \big(Z+\Up\,H_{,\bZ}\big)\, \ds F \Big|^2 \nonumber\\
 & +\big[\, 2\im\big(Z\,H_{,Z}-\bZ\,H_{,\bZ}\big)-\chi\, \big]\,\ds F \d U -2\,\d U\d V \,. \label{ContOneImpGyr_id_new2}
\end{align}
In the absence of the gyraton, ${\chi=0}$ implying ${\ds F=0}$, the metric (\ref{ContOneImpGyr_id_new2}) simplifies to the standard metric~(\ref{contipp}).

\subsection{Gyratonic Aichelburg--Sexl metric}
The expressions (\ref{OneImpGyrContTrFun_id2}), (\ref{OneImpGyrContTrFun_id1}) are valid for (\ref{chi_step}) and any metric function $H$ \emph{except the case} when
\begin{equation}
Z\,H_{,Z}-\bZ\,H_{,\bZ} = 0 \, \quad \Leftrightarrow \quad
\log\frac{Z\bZ+\Up\,\bZ H_{,\bZ}}{Z\bZ+\Up\,ZH_{,Z}} =0 \,.
\label{exceptional}
\end{equation}
This special case involves \emph{all axially symmetric geometries} with $H$ depending only on $Z\bZ$. 
In particular, this exceptional situation includes the \emph{Aichelburg--Sexl} case  ${H=-\mu \log (2\eta\bar\eta)}$, which is a famous solution to \emph{vacuum} Einstein's equations. Since ${\eta\bar\eta=Z\bar Z}$ at ${\U=U=0}$, see (\ref{trans_gyraton}), this is equivalent to
\begin{equation}
H=-\mu \log (2 Z\bar Z) \,.
\label{HAS}
\end{equation}
Instead of (\ref{OneImpGyrContTrFun_id2}), (\ref{OneImpGyrContTrFun_id1}), the functions corresponding to (\ref{HAS}) have to be taken as
\begin{equation}
 F=-\frac{\chi}{2}\,\frac{\Up}{Z\bZ-\mu\,\Up} \,, \qquad
 W=-\frac{\chi^2}{4}\,\frac{\Up}{Z\bZ-\mu\,\Up} \,,
  \label{OneImpGyrContTrFun_id_AS}
\end{equation}
and the metric is
\begin{equation}
\d s^2= 2\,\Big| \d Z +\Up\, \Big(\mu\,\frac{\d \bZ}{\bZ^2}
   +\im \frac{\chi}{2}\, \frac{\bZ \d Z+Z\d\bZ}{\bZ(Z\bZ-\mu\,\Up)} \Big)\Big|^2  -\frac{\chi^2}{2}\,\Up\, \frac{\bZ \d Z\d U +Z\d\bZ\d U }{(Z\bZ-\mu\,\Up)^2} -2\,\d U\d V \,. \label{ContOneImpGyr_id_AS}
\end{equation}
This is the new continuous metric form of the class of \emph{Frolov--Fursaev gyratons} considered in \cite{FrolovFursaev:2005}, which represent gyratonic extensions of the classic Aichelburg--Sexl solution \cite{AichelburgSexl:1971}. It describes the impulsive gravitational wave generated by a relativistic monopole point source (located at the spatial origin ${\eta=0}$ on the impulse ${\U=0}$) in which the constant $\mu$ is  related to the \emph{mass-energy of the source}, while ${\chi\, \Theta(\U)}$ determines the \emph{angular momentum density} of the gyraton, see \cite{PodolskySteinbauerSvarc:2014}.
It is convenient to introduce polar coordinates by setting ${Z=\frac{1}{\sqrt2}\,\varrho\exp(\im\phi)}$, in which the continuous metric (\ref{ContOneImpGyr_id_AS}) becomes
\begin{align}
\d s^2 =& \ \Big(1+2\mu\,\frac{\Up\, }{\varrho^2}\Big)^2\d \varrho^2
  +\Big[\Big(1-2\mu\,\frac{\Up\, }{\varrho^2}\Big)\varrho\,\d \phi
  +2\chi\frac{\Up\, }{\varrho^2}\Big(1-2\mu\,\frac{\Up\, }{\varrho^2}\Big)^{-1}\d \varrho\Big]^2
   \nonumber\\
 & \> -2\chi^2\frac{\Up\, }{\varrho^3}\Big(1-2\mu\,\frac{\Up\, }{\varrho^2}\Big)^{-2}\d \varrho\,\d U -2\,\d U\d V  \,. \label{ContOneImp_polar}
\end{align}
Notice that for ${U\le0}$ this is just ${\d s^2 = \d \varrho^2+\varrho^2\d \phi^2-2\,\d U\d V}$, which is Minkowski space in standard polar coordinates.

\emph{Without the gyratonic terms} (when the source has no spin, i.e., ${\chi=0}$), the metric (\ref{ContOneImpGyr_id_AS}) reduces to the much simpler form
\begin{equation}
\d s^2= 2\,\Big| \d Z +\mu\,\Up\, \frac{\d \bZ}{\bZ^2}  \Big|^2 -2\,\d U\d V  \,, \label{ContOneImp_id_AS}
\end{equation}
and (\ref{ContOneImp_polar}) becomes
\begin{equation}
\d s^2= \Big(1+2\mu\,\frac{\Up\, }{\varrho^2}\Big)^2\d \varrho^2
  +\Big(1-2\mu\,\frac{\Up\, }{\varrho^2}\Big)^2\varrho^2\d \phi^2 -2\,\d U\d V  \,, \label{ContOneImp_id_AS_polar}
\end{equation}
which is identical to the forms found previously in \cite{DE78, [B2]}.

\subsection{Alternative distributional and continuous forms of the gyratonic impulses}

Alternatively, instead of (\ref{imp-pp_gyrat}), it is also possible to start from the  distributional form of the {\it pp\,}-wave metric with gyratons expressed in \emph{real coordinates}
\begin{equation}\label{ro-metric}
 \d s^2 = \d\rho^2+\rho^2\,\d\varphi^2-2\,\d \U\,\d \V +2H(\rho,\varphi)\,\delta(\U)\,\d \U^2 +2\chi(\U)\,\d\varphi\, \d \U \,.
\end{equation}
This is obtained from (\ref{imp-pp_gyrat}) by the transformation
\begin{equation}\label{complex_to_real}
\eta=\frac{1}{\sqrt2}\,\rho\exp(\im\varphi)\,,\qquad\hbox{with\quad}
 J=\frac{\chi(\U)}{2\im \eta} \,,
\end{equation}
where $\chi(\U)$ is any real function. Typically, in the Einstein theory in vacuum, $H(\rho,\varphi)$ is a solution of the Laplace equation ${\triangle\, H = 0}$ (see equation (79) in Section~VII. of \cite{PodolskySteinbauerSvarc:2014}, and the discussion of more general solutions therein). The simplest explicit solution of this type is the Aichelburg--Sexl axially symmetric spacetime with ${H(\rho)=-2\mu \log \rho}$, which for ${\,\chi(\U)\not=0\,}$  describes the Frolov--Fursaev gyraton.

The corresponding continuous form (\ref{ContGenMetricSimp}) of the metric (\ref{ro-metric}) is obtained by the transformation analogous to (\ref{trans_gyraton}), namely
 \begin{eqnarray}
\U  \!\!&=&\!\! U\, , \nonumber\\
\V  \!\!&=&\!\! V+\Theta\,H+\Up\,H_{,Z}H_{,\bar Z}+W\,,  \label{trans_gyraton_alt}\\
\rho\!\!&=&\!\! \sqrt2\, \big|Z+\Up\,H_{,\bar Z}\big|\,, \nonumber\\
\varphi\!\!&=&\!\! F+\frac{1}{2\im}\log\frac{Z+\Up\,H_{,\bar Z}}{\bar Z+\Up\,H_{,Z}}\,, \nonumber
 \end{eqnarray}
in which again $H(Z,\bar Z)$ is an arbitrary real-valued function. Indeed, if the functions ${W(Z,\bar Z,U)}$, ${F(Z,\bar Z,U)}$ satisfy the equations
 \begin{equation}
F_{,U} = -\frac{\chi(\U)}  {2\, \zeta\bar\zeta}\,,   \qquad
W_{,U} = -\frac{\chi^2(\U)}{4\, \zeta\bar\zeta}\,,
 \label{condit12_alt}
 \end{equation}
where ${\zeta\bar\zeta=\eta\bar\eta}$ with ${\zeta \equiv  Z+\Up\,H_{,\bar Z}}$ defined by (\ref{zeta}), the resulting metric is exactly the continuous metric (\ref{ContGenMetricSimp}).
For the particular gyraton given by the Heaviside step function,
\begin{equation}\label{chi_step2}
\chi(\U)=\chi\, \Theta(\U)\,,\qquad\chi =\hbox{constant},
\end{equation}
(as in the example depicted in Figure~\ref{fig2}) the real functions $F$ and $W$ take the form (\ref{OneImpGyrContTrFun_id2}) and (\ref{OneImpGyrContTrFun_id1}), respectively. They are zero for all ${U\le0}$, and continuous across ${U=0}$. The corresponding continuous metric is (\ref{ContOneImpGyr_id_new2}).

\subsection{Penrose junction conditions with gyratons}
\label{sec2.4}

Using our explicit transformation (\ref{trans_gyraton}), or alternatively (\ref{trans_gyraton_alt}), we can now investigate the Penrose junction conditions (\ref{junctions}) in order to also include gyratons.

These Penrose conditions identify the corresponding points across the null hypersurface ${\cal N}$ located at ${\U=U=0}$, which separates the two halves ${{\cal M}^-(\U<0)}$ and ${{\cal M}^+(\U>0)}$ of the background Minkowski space (\ref{Mink}), see Figure~\ref{planecut}. The complete transformation relating both parts of (\ref{Mink}) to the continuous metric form (\ref{ContGenMetricSimp}) is (\ref{trans_gyraton}). Performing its limits ${U\to0^-}$ and ${U\to0^+}$, and using continuity of  the coordinates ${\{U,V,Z, \bar Z\}}$, as well as the properties of the functions $F$ and $W$, we immediately obtain
 \begin{eqnarray}
\V^+_\im  \!\!&=&\!\! \V^-_\im + H_\im \,,  \label{junction_gyraton1}\\
\eta^+_\im\!\!&=&\!\! \eta^-_\im\,, \label{junction_gyraton2}
 \end{eqnarray}
where the subscript $_\im$ indicates the value of the corresponding quantity at ${\U=U=0}$. The only discontinuity across the impulse is thus in the coordinate $\V$, in full agreement with the standard Penrose warp ${\V \to \V-H}$ across the impulsive surface prescribed by (\ref{junctions}). \emph{The presence of a gyraton thus makes no difference at all in the Penrose junction conditions} (\ref{junctions}).

Nevertheless, from the physical point of view the two distinct situations --- without a gyraton and with a gyraton --- \emph{must} have some measurable effect. It is not contained in the Penrose identification of points, but it manifests itself in the related \emph{identification of velocities} (tangent vectors to any geodesic) on both sides of the impulse. Such relations are \emph{not} part of the Penrose junction conditions (\ref{junctions}), but are contained in our generalized explicit transformation (\ref{trans_gyraton}).

Indeed, by differentiating the equations (\ref{trans_gyraton}) with respect to the parameter $\tau$ of any geodesic ${\{U(\tau),V(\tau),Z(\tau), \bar Z(\tau)\}}$ crossing ${U=0}$, for the important form of $J$ given by (\ref{chi_step}) and illustrated in  Figure~\ref{fig2}, we obtain\footnote{Due to the continuity, ${\eta=Z}$ on the impulse, and thus ${H_{\im,Z}=H_{\im,\eta}\equiv \frac{\partial}{\partial\eta}H(\eta,\bar\eta)}$, and ${H_{\im,\bar Z}=H_{\im,\bar \eta}\equiv \frac{\partial}{\partial\bar\eta}H(\eta,\bar\eta)}$.}
 \begin{eqnarray}
\dot\U^+_\im  \!\!&=&\!\! \dot\U^-_\im\,,  \label{vel_junction_gyraton0}\\
\dot\V^+_\im  \!\!&=&\!\! \dot\V^-_\im + H_{\im,Z}\,\dot\eta^-_\im + H_{\im,\bar Z}\,\dot {\bar \eta}^-_\im + \Big( H_{\im,Z}H_{\im,\bar Z} - \frac{\chi^2}{4\,\eta^-_\im\bar\eta^-_\im}\Big)\, \dot\U^-_\im\,,  \label{vel_junction_gyraton1}\\
\dot \eta^+_\im\!\!&=&\!\! \dot \eta^-_\im + \Big( H_{\im,\bar Z} -\frac{\im\chi}{2\,\bar\eta^-_\im}\Big)\, \dot\U^-_\im  \,. \label{vel_junction_gyraton2}
 \end{eqnarray}
Here we have employed that the geodesics of (\ref{ContGenMetricSimp}) are $C^1$-curves and are unique, given initial data off the impulse. This fact can be established using the Fillipov solution concept for geodesics in locally Lipschitz continuous spacetimes \cite{Steinbauer:2014}, along the lines of Remark~4.1(2) in \cite{PSSS:2015}.

For ${\chi=0}$ (implying ${F=0=W}$) we recover the conditions given by equation (4.4) in \cite{PSSS:2015}, and also in \cite{LSS:13}. The formulae (\ref{vel_junction_gyraton0})--(\ref{vel_junction_gyraton2}) extend these results by involving the gyratonic terms with $\chi$.
Clearly, there is an \emph{additional jump in the longitudinal and transverse velocities} ${\dot\V_\im}$ \emph{and} ${\dot\eta_\im}$ across the impulsive wave surface, namely
\begin{equation}
\Delta \dot\V_\im  = - \frac{\chi^2}{4\,\eta^-_\im\bar\eta^-_\im}\,\, \dot\U^-_\im\ \qquad \hbox{and}\qquad
\Delta \dot \eta_\im = -\frac{\im\chi}{2\,\bar\eta^-_\im}\,\, \dot\U^-_\im  \,. \label{jump_gyratonLV}
\end{equation}
These specific jumps become unbounded as ${\eta^-_\im \to 0}$, which is physically understandable because the singular gyratonic source is located at the spatial origin ${\eta=0}$ of the impulsive surface. It is also interesting to observe that
\begin{equation}
|\Delta \dot \eta_\im|^2 = -\,\dot\U^-_\im \,\Delta \dot\V_\im \,. \label{jump_gyratonLVcorrel}
\end{equation}
This identity is related to the conservation of normalization of the four-velocity, which is guaranteed by the $C^1$-regularity of geodesics and the continuity of the metric.

The effect of the gyraton on test particles moving along geodesics is even more nicely seen if we employ real polar coordinates and the explicit transformation (\ref{trans_gyraton_alt}). This yields
\begin{eqnarray}
\V^+_\im = \V^-_\im + H_\im \,,\qquad
\rho^+_\im = \rho^-_\im \,,\qquad \varphi^+_\im = \varphi^-_\im \,,
\label{polar_junction_gyraton}
\end{eqnarray}
so that both the radial position $\rho_\im$ and the polar position $\varphi_\im $ are  continuous (uneffected by the presence of a gyraton), but \emph {there is a discontinuity in the corresponding velocities}
 \begin{eqnarray}
\dot\V^+_\im  \!\!&=&\!\! \dot\V^-_\im
+ \frac{1}{\sqrt 2}\Big(e^{\im\varphi^-_\im} H_{\im,Z} + e^{-\im\varphi^-_\im} H_{\im,\bar Z}\Big)\,\dot\rho^-_\im
+ \frac{\im}{\sqrt 2}\Big(e^{\im\varphi^-_\im} H_{\im,Z} - e^{-\im\varphi^-_\im} H_{\im,\bar Z}\Big)\,\rho^-_\im \dot\varphi^-_\im \nonumber\\
&& + \Big( H_{\im,Z}H_{\im,\bar Z} - \frac{\chi^2}{2(\rho^-_\im)^2}\Big)\, \dot\U^-_\im\,,  \label{polar_junction_gyraton1}\\
\dot \rho^+_\im\!\!&=&\!\! \dot \rho^-_\im + \frac{1}{\rho^-_\im}\Big( H_{\im,Z} + H_{\im,\bar Z} \Big)\, \dot\U^-_\im  \,, \label{polar_junction_gyraton2}\\
\dot \varphi^+_\im\!\!&=&\!\! \dot \varphi^-_\im + \bigg[\frac{\im}{\sqrt{2}\,\rho^-_\im} \Big(e^{\im\varphi^-_\im} H_{\im,Z} - e^{-\im\varphi^-_\im} H_{\im,\bar Z}\Big) -\frac{\chi}{(\rho^-_\im)^2}\bigg]\, \dot\U^-_\im  \,. \label{polar_junction_gyraton3}
 \end{eqnarray}
In particular, the gyraton has \emph{no effect on the radial velocity $\dot \rho_\im$}, but it causes a \emph{specific jump of the axial velocity $\dot \varphi_\im$} given by the term ${\,\chi\,/(\rho^-_\im)^{2}}$. This is consistent with the physical interpretation of the gyratonic term: it encodes the additional rotational (spin) character of the source of the impulsive gravitational wave. The gyraton also affects the longitudinal velocity $\dot\V_\im$ via the term ${\,\chi^2\,/2(\rho^-_\im)^{2}}$ in the expression (\ref{polar_junction_gyraton1}). It agrees with the studies presented in \cite{FrolovFursaev:2005, YoshinoZelnikovFrolov:2007, SSS:2016}.

\section{Extension to any cosmological constant $\Lambda$}
\label{Sec3}

These results on impulsive {\it pp\,}-waves with gyratons can be generalized to any background of constant curvature, i.e., to gyratonic impulses propagating in \emph{de~Sitter} or \emph{anti-de~Siter} spacetimes.

Indeed, it was already shown in \cite{PodolskyGriffiths:1999a} that the original Penrose ``cut and paste'' construction method in Minkowski space (summarized here in Subsection~\ref{subs1}) can be extended to any $\Lambda$ by applying exactly \emph{the same junction conditions} (\ref{junctions}) to a more general background metric generalizing (\ref{Mink}), namely
 \begin{equation}
\d s_0^2= \frac{2\,\d\eta\,\d\bar\eta -2\,\d\U\,\d\V}
{[\,1+\frac{1}{6}\Lambda(\eta\bar\eta-\U\V)\,]^2}\,.
 \label{conf*}
 \end{equation}
This introduces impulsive waves in the de Sitter (${\Lambda>0}$) or anti-de~Sitter ($\Lambda<0$)
universes.

Of course, the geometry of such impulses depends on $\Lambda$ since
the null hypersurface ${\cal N}$, given by ${\,\U=0}$, along which the spacetime is cut
into the two halves ${\cal M}^-$ and ${\cal M}^+$ and re-attached with a specific ``warp'' (\ref{junctions}), has the induced 2-metric
${\d\sigma^2= 2\,(\,1+\frac{1}{6}\,\Lambda\,\eta\,\bar\eta\,)^{-2}\,\d\eta\,\d\bar\eta}$ with the Gaussian curvature ${K=\frac{1}{3}\Lambda}$. Thus, for ${\Lambda=0}$ the impulsive wave surface is a \emph{plane}, for ${\Lambda>0}$ it is a \emph{sphere}, while for ${\Lambda<0}$ it is a \emph{hyperboloid}. The geometry of these nonexpanding impulsive spherical and hyperboloidal waves was described in detail in~\cite{[B5]} using various coordinate representations.

It was also shown in \cite{PodolskyGriffiths:1999a} that applying the same transformation
(\ref{trans}) to the metric (\ref{conf*}), the \emph{explicit continuous form} of the impulsive metric is obtained, namely
 \begin{equation}
\d s^2= \frac{2\,|\d Z+\Up(H_{,\bar ZZ}\d Z+H_{,\bar Z\bar Z}\d\bar Z)|^2-2\,\d U\d V}
{[\,1+\frac{1}{6}\Lambda(Z\bar Z-UV-\Up G)\,]^2}\ ,
 \label{conti}
 \end{equation}
where ${G(Z,\bar Z)\equiv H-ZH_{,Z}-\bar ZH_{,\bar Z}}$.

Notice that this metric is conformal to the continuous form of impulsive {\it pp\,}-waves (\ref{contipp}), to which it reduces in the case ${\Lambda=0}$. Although it is continuous across the null hypersurface ${U=0}$, the discontinuity in the derivatives of the metric
introduces  impulsive components in the Weyl and curvature tensors proportional
to the Dirac  distribution \cite{PodolskyGriffiths:1999a}. The metric (\ref{conti}) thus explicitly describes  impulsive waves in de~Sitter, anti-de~Sitter or Minkowski backgrounds.

Moreover, \emph{for any} ${\Lambda}$ the transformation (\ref{trans}) automatically incorporates the Penrose junction conditions (\ref{junctions}) for reattaching the two halves of the spacetime ${\cal M}^-$ and ${\cal M}^+$ with the warp ${\V \to \V-H}$.

The corresponding \emph{distributional form} of these impulsive solutions reads
 \begin{equation}
\d s^2= \frac{2\,\d\eta\,\d\bar\eta-2\,\d \U\,\d \V +2H(\eta,\bar\eta)\,\delta(\U)\,\d \U^2}
{[\,1+\frac{1}{6}\Lambda(\eta\bar\eta-\U\V)\,]^2}\,.
 \label{confppimp}
 \end{equation}
It is obtained from the continuous form of the impulsive-wave metric (\ref{conti}) by
applying the transformation (\ref{trans}) if the distributional terms arising from the derivatives of $\Theta(\U)$ and $\Up(\U)$ are also kept (see  \cite{PodolskyGriffiths:1999a}, \cite{HorItz99} for more details). The impulse, located on the wavefront ${\U=0}$, propagates in a background spacetime of constant curvature (\ref{conf*}). For ${\Lambda=0}$ this is the Brinkmann form of an impulsive
{\it pp\,}-wave (\ref{ppimp}) in Minkowski space, while for ${\Lambda\not=0}$ it represents an impulse propagating in curved (anti-)de~Sitter universe (the metric (\ref{confppimp}) is conformal to the Brinkmann metric).

In fact, the family of impulsive spacetimes (\ref{confppimp}) contains \emph{all nonexpanding impulses} in Minkowski, de~Sitter or anti-de~Sitter universes that can
be constructed from the \emph{whole Kundt class} of type~N solutions with a cosmological constant $\Lambda$ \cite{Kundt, KSMH, GP:2009, ORR, [A1]}. A systematic analysis of such distributional limits of Kundt sandwich waves was performed in \cite{[B1]}, for a  review and explicit transformations see \cite{PSSS:2015}. This recent review also summarizes the most important explicit impulses of this type, namely the axially symmetric Hotta--Tanaka solution \cite{HotTan93}, which was obtained by boosting the Schwarzschild--de~Sitter or  Schwarzschild--anti-de~Sitter black hole to the speed of light (the analogue of the Aichelburg--Sexl solution \cite{AichelburgSexl:1971} for ${\Lambda\not=0}$), and more general nonexpanding impulsive waves generated by null multipole particles \cite{[B4]}. Such solutions are more clearly expressed and analyzed by employing the 5-dimensional representation of the (anti-)de~Sitter spacetime~\cite{[B5], [B6], PO:01, SSLP:2016, SS:2017}.

\subsection{Continous form of impulsive waves with gyratons and $\Lambda$}

Interestingly, an explicitly continuous metric form of gyratonic impulsive waves in de~Siter and anti-de~Sitter spacetimes are obtained by applying the discontinuous transformation (\ref{trans_gyraton})--(\ref{condit1}) on the background metric (\ref{conf*}), which yields
\begin{equation}
 \d s^2= \frac{2\,\big| \ds \zeta +\im \zeta\, \ds F \big|^2
  +2\big[ \im\Theta \big( \zeta H_{,Z}-
  \bzeta H_{,\bar Z} \big) \ds F-\ds W \big]\d U -2\,\d U\d V }
  {[\,1+\frac{1}{6}\Lambda(Z\bar Z-UV-\Up G)\,]^2}\,,
 \label{contigyrL}
\end{equation}
where
\begin{equation}
G(Z,\bar Z,U)\equiv H-ZH_{,Z}-\bar ZH_{,\bar Z}+W \,.
\label{Ggeneral}
\end{equation}
It is obvious that for ${\Lambda=0}$ this reduces to (\ref{ContGenMetricSimp}), while in the absence of gyratons (${W=0=F}$ implying ${J=0}$) one recovers the metric form (\ref{conti}). If both ${\Lambda=0}$ and ${J=0}$, the classical metric form (\ref{contipp}) is obtained.

Indeed, to establish this, note that the \emph{numerator} of (\ref{contigyrL}) is identical to (\ref{ContGenMetricSimp}). As demonstrated in Section~\ref{Sec2}, this is equivalent to the distributional metric (\ref{imp-pp_gyrat}) via the transformations (\ref{trans_gyraton})--(\ref{condit1}), i.e., to the background (\ref{Mink}) for ${\U\equiv U\not=0}$. The (anti-)de~Sitter background (\ref{conf*}) involves the additional conformal factor, namely the square of ${\,1+\frac{1}{6}\Lambda(\eta\bar\eta-\U\V)}$ in the \emph{denominator}. Performing the transformation (\ref{trans_gyraton}) and using the multiplication rules ${\,\Up=U\Theta\,}$ and ${\,\Theta^2=\Theta\,}$, this expression becomes
\begin{equation}
1+{\textstyle\frac{1}{6}}\Lambda\Big(Z\bar Z-UV-\Up (H-ZH_{,Z}-\bar ZH_{,\bar Z})-U W \Big) \,.
\end{equation}
Now, the key observation is that ${\,UW\equiv \Up W\,}$ whenever ${W=0}$ for all ${U\le 0}$, as is the case in the explicit examples (\ref{OneImpGyrContTrFun_id1}) and (\ref{OneImpGyrContTrFun_id_AS}). Consequently, ${1+\frac{1}{6}\Lambda(\eta\bar\eta-\U\V)=1+\frac{1}{6}\Lambda (Z\bar Z-UV-\Up G) }$, where $G$ is defined by (\ref{Ggeneral}). This gives the denominator in (\ref{contigyrL}), completing the argument.

\subsection{Distributional form of impulsive waves with gyratons and $\Lambda$}

Combining the results and relations employed in the previous subsection, we may also conclude that the class of spacetimes studied here can be written in the following form
 \begin{equation}
\d s^2= \frac{2\,\d\eta\,\d\bar\eta-2\,\d \U\,\d \V +2H(\eta,\bar\eta)\,\delta(\U)\,\d \U^2
+2J(\eta,\bar\eta,\U)\,\d\eta\,\d \U + 2\bar J(\eta,\bar\eta,\U)\,\d\bar\eta\,\d \U}
{[\,1+\frac{1}{6}\Lambda(\eta\bar\eta-\U\V)\,]^2}\,.
 \label{impgyrL}
 \end{equation}
This is a combination of the metric (\ref{imp-pp_gyrat}) for impulsive {\it pp\,}-waves with
the gyratonic off-diagonal terms (generalized Brinkmann metric\cite{Brinkmann:1925}) in the numerator, with the conformal factor ${[\,1+\frac{1}{6}\Lambda(\eta\bar\eta-\U\V)\,]^{-2}}$. This \emph{distributional form} of the metric is a \emph{general expression for impulsive gyratonic  waves propagating in any spacetime of constant curvature}, that is Minkowski, de~Sitter or anti--de~Sitter universe, according to the sign of the cosmological constant $\Lambda$. The metric (\ref{impgyrL}) is clearly conformal to the distributional metric (\ref{imp-pp_gyrat}) for impulsive gyratonic {\it pp\,}-waves.

\subsection{Penrose junction conditions with gyratons and $\Lambda$}

Similarly as in Subsection~\ref{sec2.4}, we can now obtain and discuss the extended Penrose junction conditions in the presence of gyratons when ${\Lambda\not=0}$. They identify the corresponding points across the impulsive null hypersurface ${\cal N}$ located at ${\U=U=0}$, which separates ${{\cal M}^-(\U<0)}$ and ${{\cal M}^+(\U>0)}$ of the background (anti-)de~Sitter manifold (\ref{conf*}).

The full transformation relating both parts ${{\cal M}^-}$ and ${{\cal M}^+}$ in the form (\ref{conf*}) to the continuous metric form (\ref{contigyrL}) is \emph{again given by} (\ref{trans_gyraton}). Interestingly, the inclusion of the cosmological constant $\Lambda$ only occurs via the conformal factor ${[\,1+\frac{1}{6}\Lambda(\eta\bar\eta-\U\V)\,]^{-2}=[\,1+\frac{1}{6}\Lambda (Z\bar Z-UV-\Up G)\,]^{-2}}$ in (\ref{contigyrL}), where $G$ is defined by (\ref{Ggeneral}). Moreover, evaluated on the impulse it takes the very simple form
\begin{equation}
 [\,1+{\textstyle \frac{1}{6}\Lambda}\,\eta\bar\eta\,]^{-2}=[\,1+{\textstyle \frac{1}{6}\Lambda}\, Z\bar Z\,]^{-2}.
 \label{confac}
\end{equation}

Performing the limits ${U\to0^-}$ and ${U\to0^+}$ of the transformation (\ref{trans_gyraton}), employing the continuity of the coordinates ${\{U,V,Z, \bar Z\}}$, we obtain
 \begin{eqnarray}
\V^+_\im  \!\!&=&\!\! \V^-_\im + H_\im \,,  \label{junction_gyraton1L}\\
\eta^+_\im\!\!&=&\!\! \eta^-_\im\,. \label{junction_gyraton2L}
 \end{eqnarray}
 Clearly, these are the same as the relations (\ref{junction_gyraton1}), (\ref{junction_gyraton2}). We have thus recovered  the Penrose junction conditions (\ref{junctions}) for proper identification of points (positions) across the impulse which are valid for any value of the cosmological constant, see \cite{PodolskyGriffiths:1999a, Pod2002b, PSSS:2015}. Of course, for ${\Lambda\not=0}$ the coordinates ${\{\U, \V, \eta, \bar \eta\}}$ are not the usual Minkowski double null coordinates, but the  coordinates of the (anti-)de~Sitter conformally flat metric (\ref{conf*}).

These gyratonic impulsive waves propagating in the (anti-)de~Sitter universe have a specific effect on the \emph{velocities} of free test particles. As in the case ${\Lambda=0}$, the corresponding junction conditions are obtained by differentiating the explicit transformation (\ref{trans_gyraton}) with respect to the parameter $\tau$ of any geodesic, and comparing its limits  ${U\to0^-}$ and ${U\to0^+}$. Due to (\ref{confac}), we obtain the same relations (\ref{vel_junction_gyraton0}), (\ref{vel_junction_gyraton1}), (\ref{vel_junction_gyraton2}) as in the Minkowski background case.

\section{Conclusions}

Let us summarize the main results of our work:

\begin{itemize}

\item
We have derived the new \emph{continuous metric form} (\ref{ContGenMetricSimp}) for impulsive  {\it pp\,}-waves \emph{with gyratons}, which naturally generalizes the classical metric (\ref{contipp}) without the gyratonic terms.

\item
We have found the \emph{transformation} (\ref{trans_gyraton})--(\ref{condit1}) relating the continuous metric form (\ref{ContGenMetricSimp}) to the \emph{distributional metric form} (\ref{imp-pp_gyrat}), which --- compared to (\ref{imp-pp}) --- contains the additional off-diagonal metric function $J$ representing the gyraton. This is an extension of the classical transformation~(\ref{trans}).

\item
We have explicitly presented the continuous form (\ref{ContOneImpGyr_id_new2}) of the impulsive metric for the \emph{key example} (\ref{chi_step}) when  $J$ is proportional to the Heaviside \emph{step function} of retarded time, \emph{including also the exceptional case}  (\ref{ContOneImpGyr_id_AS}), or (\ref{ContOneImp_polar}), of the Frolov--Fursaev gyratons constructed from the axially symmetric Aichelburg--Sexl vacuum solution.

\item
We have proved that the \emph{Penrose junction conditions} (\ref{junctions}) for identifying the corresponding \emph{points} in the ``cut and paste'' construction method \emph{remain valid even in the presence of gyratons} (because the additional continuous functions $W$ and $F$ vanish on the impulsive hypersurface).

\item
Nevertheless, the \emph{presence of a gyraton manifests itself through the ``derivatives'' of the junction conditions}, as represented by  the transformation (\ref{trans_gyraton})--(\ref{condit1}). This leads to a \emph{specific jump in the velocities} (\ref{vel_junction_gyraton1}), (\ref{vel_junction_gyraton2}) of test particles crossing the gyrating impulse. As clearly demonstrated by (\ref{polar_junction_gyraton2}), (\ref{polar_junction_gyraton3}), \emph{the gyraton affects only the axial component of the transverse velocity}.

\item
Finally, we have generalized all these new results (obtained for the case of impulsive {\it pp\,}-waves with gyratons, and thus necessarily when ${\Lambda=0}$) to \emph{any value of the cosmological constant} ${\Lambda}$.
In particular, starting from the unified, conformally flat, form of the  metric (\ref{conf*}) representing the de Sitter (${\Lambda>0}$), anti-de~Sitter ($\Lambda<0$) or flat Minkowski  ($\Lambda=0$) backgrounds, we demonstrated that the \emph{transformation} (\ref{trans_gyraton})--(\ref{condit1}) \emph{can by applied for any} $\Lambda$, and it relates the new \emph{continuous metric form} (\ref{contigyrL}), (\ref{Ggeneral}) to the corresponding \emph{distributional metric form} (\ref{impgyrL}) of gyratonic impulsive waves propagating in de~Sitter or anti-de~Sitter spacetimes.

\end{itemize}

There are also several possibilities for further extension of the above results. For example, it would be interesting to find the continuous metric form for gyratonic impulsive waves with more general profiles of energy and angular momentum. These are going to be topics of subsequent studies.

\section*{Acknowledgements}

JP and R\v{S} were supported by the Czech Science Foundation grant GA\v{C}R 17-01625S.
RS and CS acknowledge the support of FWF grants P25326 and P28770.

\end{document}